# A Separable Temporal Convolution Neural Network with Attention for Small-Footprint Keyword Spotting

*Shenghua Hu[1], Jing Wang[1], Yujun Wang[2], Lidong Yang[3], Wenjing Yang[1]*

[1]School of Information and Electronics, Beijing Institute of Technology, Beijing, China
[2]Xiaomi Inc., Beijing, China
[3]School of Information Engineering, Inner Mongolia University of Science and Technology, Baotou, China

```
hushenghua@bit.edu.cn, wangjing@bit.edu.cn, wangyujun@xiaomi.com,
       yldnkd@imust.edu.cn, yangwenjing3@xiaomi.com
```

## Abstract

Keyword spotting (KWS) on mobile devices generally requires a small memory footprint. However, most current models still maintain a large number of parameters in order to ensure good performance. To solve this problem, this paper proposes a separable temporal convolution neural network with attention, it has a small number of parameters. Through the time convolution combined with attention mechanism, a small number of parameters model (32.2K) is implemented while maintaining high performance. The proposed model achieves 95.7% accuracy on the Google Speech Commands dataset, which is close to the performance of Res15(239K), the state-of-the-art model in KWS at present.

**Index Terms**: keyword spotting, convolutional neural network, temporal convolution, attention mechanism, mobile device

## 1. Introduction

KWS aims at detecting predefined keywords in audio signals and is widely implemented in hands-free control of mobile devices. Recently, many human-machine interfaces (HCI) rely on KWS, such as Apple Siri [1], Microsoft Cortana, Amazon Alexa [2,3] and Google Assistant [4], etc. These systems exploit powerful neural network models, which usually run in the cloud. Lightweight KWS models enable the development of many novel engineering applications, such as high-real-time voice control equipment and operation without Internet coverage. However, it's still a very challenging task to implement a fast and accurate KWS model on mobile devices with limited hardware resources.

Nowadays, there are two mainstreams to realize KWS, one based on large vocabulary continuous speech recognition (LVCSR) [5] and another based on hidden Markov model (HMMS) [6]. The main weaknesses in these methods are huge memory cost and calculations, which makes it impossible to be applied on mobile devices with limited hardware resources.

With the success of deep learning in various fields, KWS based on neural networks has become the mainstream at present [7,8,9,10]. KWS based on Convolutional Neural Network (CNN) has achieved high accuracy results [11]. The ResNet-based KWS system proposed by Tang and Lin achieved 95.8% accuracy on the Google Speech Commands V1 [7]. This structure applied residual connections to deepen the number of layers of the network, but with too many hidden layers and filters, it still has 238K parameters. The Att-RNN proposed by Andrade et al uses Attention instead of Average pooling to improves accuracy，but the traditional CNN and Bi-directional Long Short-Term Memory (BLSTM) in Att-RNN are too bulky (202K) [7,12,13].

Depthwise separable convolution (DSCNN) decomposes traditional convolution operations into depthwise (DW) and pointwise (PW), greatly reducing the number of model parameters and calculations with almost no performance loss. In current many tasks, it was successful [14]. Similarly, it is also well applied to KWS [8]. However, DSCNN needs enough hidden layers to obtain the receptive field, which will also lead to an increase in the number of parameters and calculations.

In this paper we propose a separable temporal convolution neural network with attention for keyword spotting, denoted as ST-Conv. The proposed model takes MFCC features as input and uses depthwise separable temporal convolution to capture local features. Besides, we use Bidirectional Gated Recurrent Unit (BGRU) and Attention to capture global features. The proposed model takes the advantages of the above structure with a drop of the number of parameters and calculations for KWS while maintaining accuracy. Our contributions are as follows:

- We proposed a novel neural network structure based on CNN and attention mechanism for small-footprint KWS on mobile devices, which can achieve the current state-of-the-art performance. In addition, we analyzed the role of each layer in the proposed neural network.
- We conducted a series of comparative experiments with the current state-of-the-art model on the publicly available data sets Google Speech Commands V1 [15]. The experimental results show that compared with the state-of-the-art ResNet-15, our proposed model achieves a nearly consistent accuracy rate (95.7% vs 95.8%) and a much smaller number of model parameters (32.2K vs 239K).

The rest of this paper is organized as follows. Section 2 describes the proposed models. Section 3 introduces the experimental setup and results. Section 4 concludes the paper.

## 2. Network Architecture

### 2.1. Separable temporal convolution

Figure 1 uses the first two layers of our proposed network as an example to illustrate the separable temporal convolution using MFCC as input. There is no padding in the first layer, and the given input $\in \mathbb{R}^{99\times40}$, filter $\in \mathbb{R}^{1\times40\times40}$. The output of the 2-dimensional convolution is the first layer output $\in \mathbb{R}^{99\times1\times40}$. In the second layer, zero padding is applied to match the input and output resolution, given filter $\in \mathbb{R}^{3\times1\times40}$ and perform depthwise separable convolution to obtain the output of the second layer $\in \mathbb{R}^{99\times1\times40}$. Note that the dimensions of first layer and second layer outputs are the same.

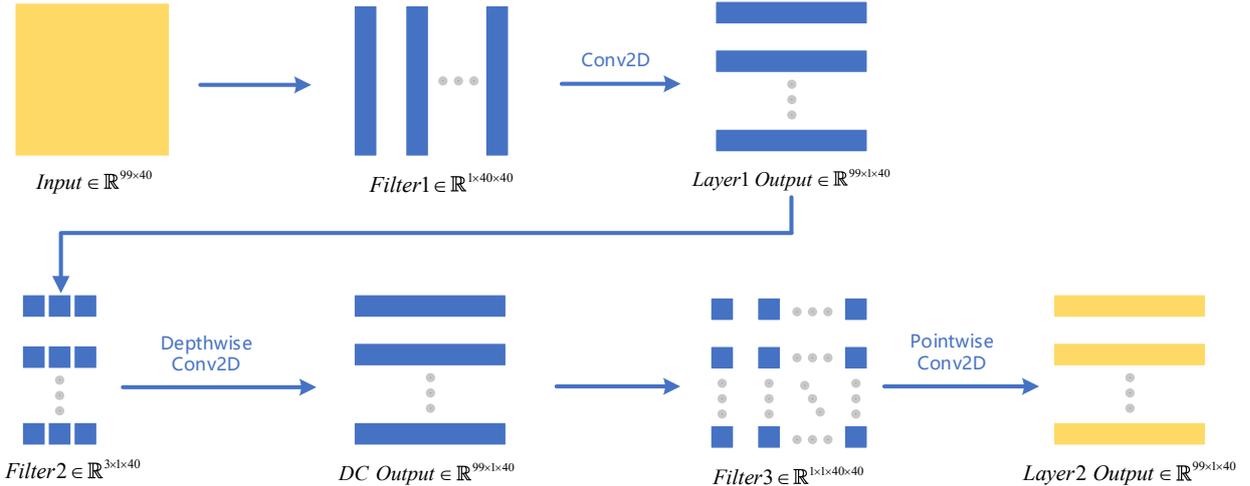

Figure 1: *The first two layers of the proposed model illustrate the separable temporal convolution. The input of the network is a 40-dimensional MFCC feature extracted from audio with a duration of 1 second. The first layer is a 2-dimensional convolution with a special convolution kernel, no padding and stride is set to 1. Through this convolution method, the feature size in each subsequent channel can be reduced to 1. Starting from the second layer, we use separable convolution, by using the zero padding, set stride to 1 to ensure that the output and input have the same size.*

CNN is a classic neural network structure, which can extract high-level features from low-level features. However, the current CNN structure usually uses a small-sized convolution kernel (in most cases, the convolution kernel size is $3\times3$). In case of KWS, we need a large number of stacked convolutional layers to obtain a larger receptive field. Although We can alleviate this problem by increasing the stride or dilated convolution, etc, we adopt these methods, most of the current models which still have a large number of parameters and high computational complexity are difficult to run on mobile devices.

In order to achieve a model with both lightweight and high accuracy, we proposed the Separable Temporal Convolution. Our main idea is to extract the information on the frequency domain of MFCC first, and then expand the receptive field in the time dimension gradually. Therefore, we use a convolution kernel $\in \mathbb{R}^{1\times40\times40}$ in the first layer of the network. Our input feature is a 40-dimensional MFCC, so through the first layer of convolution we can ensure that each subsequent layer can utilize all frequency domain information. In order to maximize the receptive field of the neural network in the limited layers, we use dilated convolution in the second and subsequent layers. Here we apply an exponential sizing schedule [16], for the i-th layer, its dilation rate is $d = 2^{\lfloor \frac{i}{3} \rfloor}$.

According to our experiment results, our method has the following advantages over traditional CNN:

**More suitable for small-footprint voice sequence tasks.** The traditional 2D convolution will extract 2 dimensions of makes it an irreplaceable advantage in image processing tasks. However, in the small-footprint speech sequence task, extracting two dimensions of information at the same time may not be an excellent strategy, because it makes the cost of expanding receptive field unacceptable. ST-Conv considers the voice signal as sequence information, and preferentially extracts the frequency domain information of each frame, which is also a method that conforms to human intuition.

**Lower amount of parameter and calculation.** Using filter $\in \mathbb{R}^{1\times40\times40}$ in the first layer can not only extract the information in the entire frequency domain, but also reduce the size of the convolution kernel of the subsequent layer from $3\times3$ to $3\times1$, reducing the number of parameters by two-thirds in subsequent layer. In order to further reduce calculation of the network, we adopt a separable convolution method at the same time which can significantly reduce the calculation when the number of channels is large.

**Expand the receptive field more efficiently.** Starting from the second layer，we apply the dilated convolution with a more efficient exponential sizing schedule [16]. Therefore, the total receptive field of our proposed model reached 121 in the CNN stage.

### 2.2. Model Architecture

In the convolutional network part of the model, we refer to the ResNet proposed by He et al [17]. He et al believed that learning residuals might be easier than learning the original mapping of deep convolutional neural networks. We designed the block depicted in Figure 2a for the neural network, in which a separable convolution with the convolution kernel size of $3\times1$ is used, with zero padding, channel size c=40 and dilation rate $d = 2^{\lfloor \frac{i}{3} \rfloor}$. After the convolutional layer, there are ReLU activation units and a batch normalization layer and residuals were added between the blocks [18].

Figure 2b is the overall schematic diagram of our network. The 40-dimensional MFCC obtains 99×1×40 output after 13-layer CNN calculation. At this time, we squeeze the second dimension to reduce the output to 2-dimensional, and then send the output into bidirectional gated recurrent unit to capture long-term dependence in speech. After that is a critical step of our structure, we refer to the Att-RNN [9] proposed by Andrade et al and the Shared Weight Self-Attention [19] proposed by Bai et al. As shown in Figure 2c, we extract an output vector *x* of the BGRU layer, project *x* with a matrix *w* to obtain *q*, which is the query in the attention mechanism. Then we use the same matrix *w* to project the entire output of the BGRU to obtain *u*, which is key and value in attention mechanism. Finally, we use query to determine which parts of the audio are most relevant. We chose the 49th vector output by BGRU to project the query. Theoretically, this choice is arbitrary. After multiple layers of stacked CNN and BGRU, any vector should have enough "memory". Finally, we send the output of SWSA to the fully connected layer and Softmax for classification to get the final output of the network.

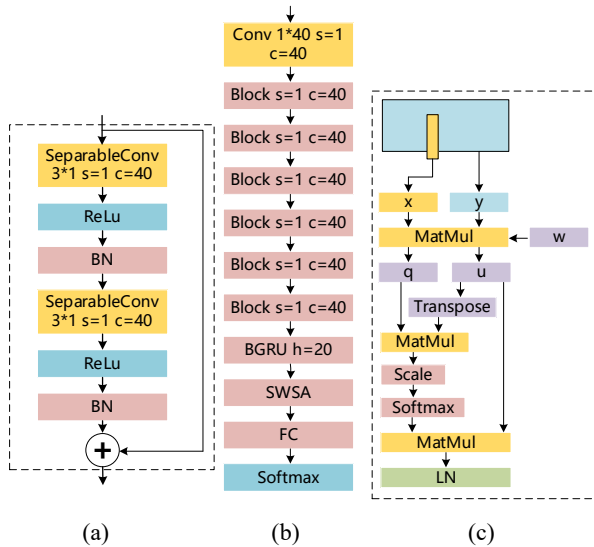

Figure 2: *(a) is the building block (denoted as Block) of ST-Conv. (b) is a schematic of the architecture. (c) is explanatory diagram of SWSA. BN and LN denote batch normalization and layer normalization. BGRU, SWSA and FC denote bidirectional gated recurrent unit, fully connected layer and shared weight self-attention respectively. Note that 's' and 'c' indicates stride and channel size.*

Table 1: *Classification of keywords and fillers.*

| Keywords | "down", "go", "left", "no", "off", "on", "right", "stop", "up", "yes" |
|---|---|
| Fillers | "bed", "bird", "cat', "dog", "happy", "house", "marvin", "sheila", "tree", "wow", "zero", "one", "two", "three", "four", "five", "six", "seven", "eight", "nine" |

Table 2: *The configuration of ST-Conv. t and f are the size of the convolution kernel in time and feature dimensions, respectively. n is the dimension of the output feature. d represents the dilation rate. Par is the number of parameters of the weight matrix. Mult is the number of multipliers in matrix multiplication.*

| Type | t | f | n | d | Par. | Mult. |
|---|---|---|---|---|---|---|
| Conv | 1 | 40 | 40 | 1 | 1.6K | 158.4K |
| Block*6 | 3 | 1 | 40 | $2^{\lfloor \frac{i}{3} \rfloor}$ | 20.6K | 2043.3K |
| BGRU | - | - | 40 | - | 7.4K | 724.6K |
| SWSA | - | - | 40 | - | 1.6K | 167.9K |
| FC | - | - | 20 | - | 800 | 0.8K |
| Softmax | - | - | 11 | - | 220 | 0.22K |
| Total | - | - | - | - | 32.2K | 3.09M |

## 3. Experiments

### 3.1. Datasets

We use the Google Speech Commands dataset to evaluate our proposed model and baselines. For a fairer comparison, we used the same version (V1) as the baselines. The dataset consists of 64,752 recordings, containing a total of 30 words. Each recording is 1 second in duration and contains one word. 10 words are used as keywords, and the remaining 20 words are used as fillers. The words specifically divided into keywords and fillers are listed in Table 1. We use the standard list provided by the data set to divide the data set into training set, development set and test set. The specific situation is that 51088 voices are divided into the training set, 6798 voices are divided into the development set, and 6,835 voices are divided into the test set.

### 3.2. Experimental setup

We extract 40-dimensional MFCC features from each piece of audio as the input of the neural network, the frame length is 25ms and the frame shift is 10ms. Each piece of audio is divided into 99 frames in the time dimension, then we stack and send them to the neural network.

We give the configuration of each layer of the proposed neural network in Table 2, and also give the number of parameters and the number of multiplication operations for each layer. We use a multi-head attention mechanism at the SWSA layer, with the number of heads as 4. We use tensorflow to train and evaluate our model [20]. Cross entropy is employed as the loss function. The initial learning rate is set to 0.001. After each epoch, the model will be evaluated on the development set. If the loss is not observed to decrease significantly (3%) compared to the previous epoch, then we will reduce the learning rate to 60%. In addition, when learning rate lower than $1\times10^{-5}$, it will be limited to $1\times10^{-5}$. At the same time, we force each learning rate to maintain at least 2 epochs. We use Adam algorithm as the optimizer [21]. The total number of epochs is set to 80. The mini-batch size is 32. We use the early stopping strategy, which means that we save the best performing model on the development set as the final model [22].

We use accuracy as the most important indicator to evaluate the performance of the model. We train each model 5 times, and then we evaluated the results with 95% confidence intervals. Receiver operating characteristic (ROC) curves are also used to evaluate our model. Its x-axis is the false alarm rate, and the y-axis is the false reject rate. We scan in the threshold [0.0, 1.0], calculate the curve of a specific keyword, and then average vertically to generate the overall curve of the specific model. The smaller the area under the curve, the better the model.

Table 3: *Comparison of the baseline models and the proposed models. We list the test accuracy, the number of parameters and multipliers. All of the baselines have been reimplemented by us.*

| Model | Test accuracy | Par. | Mult. |
| --- | --- | --- | --- |
| Res8 | 94.2% ± 0.341 | 111K | 30M |
| Res8-Narrow | 90.3% ± 0.877 | 20K | 5.65M |
| Res15 | 95.8% ± 0.256 | 239K | 894M |
| Res15-Narrow | 94.1% ± 0.535 | 43K | 160M |
| Att-RNN | 95.6% ± 0.213 | 202K | 22.6M |
| TC-ResNet8 | 95.4% ± 0.412 | 66K | 1.63M |
| TC-ResNet14 | 95.7% ± 0.324 | 137K | 3.26M |
| ST-Conv | 95.7% ± 0.295 | 31K | 3.09M |
| ST-Conv-Narrow | 94.3% ± 0.512 | 9.4K | 0.67M |
| ST-Conv-Avg | 94.9% ± 0.356 | 30K | 2.92M |

### 3.3. Baselines

We have selected the model structure listed below as the baseline to verify the advantages of our proposed model in terms of parameter quantity and accuracy.

**Res8, Res8-Narrow, Res15, and Res15-Narrow** [7]. Res15 is the current state-of-the-art model structure on the speech commands data set. Several other models are its variants. Narrow means there is a reduction on the output channel.

**Att-RNN** [9]. It is also the current state-of-the-art model structure on the speech commands data set.

**TC-ResNet8, TC-ResNet14** [10]. They are composed of a series of one-dimensional temporal convolution residual blocks. The number after TC-ResNet represents the number of layers of the network.

### 3.4. Experimental results

The comparison between our proposed model and all baselines is shown in Table 3. All are obtained on the standard test set of Speech Commands V1. In order to verify the effect of further reducing the number of parameters on accuracy, we designed ST-Conv-Narrow, a variant of ST-Conv. ST-Conv-Narrow is obtained by reducing the number of channels in the convolution process to 20 based on the structure of ST-Conv. At the same time, we reduce the output dimension of BGRU to 20. From the experimental results, compared with ST-Conv, the parameter of ST-Conv-Narrow is reduced by two-thirds and the accuracy rate is reduced by 1.5%. We think this is acceptable.

In addition, in order to verify the effectiveness of the SWSA layer in ST-Conv, we design another variant of ST-Conv, ST-Conv-Avg. ST-Conv-Avg is obtained by replacing the SWSA layer in ST-Conv with the Average Pooling layer. Compared with ST-Conv, ST-Conv-Avg using Average Pooling has a 0.8% drop in accuracy. So SWSA is verified to be effective in ST-Conv.

As for the baseline in our experiments, Tang and Lin proposed a variety of variants based on the ResNet structure. Res15 is a variant with the best performance. Res8 reduces the number of CNN layers in the neural network. Res15-Narrow and Res8-Narrow further reduce the number of channels. Shown in Table 3, Res15 is indeed one of the best performing models. But its number of parameters and multipliers are too large. Res8-Narrow is the most compact variant they proposed, with a very small model size. However, it paid too high a price in terms of accuracy in order to reduce the number of parameters. Res15-Narrow has a slightly higher number of parameters than ST-Conv, but it is far less accurate than ST-Conv.

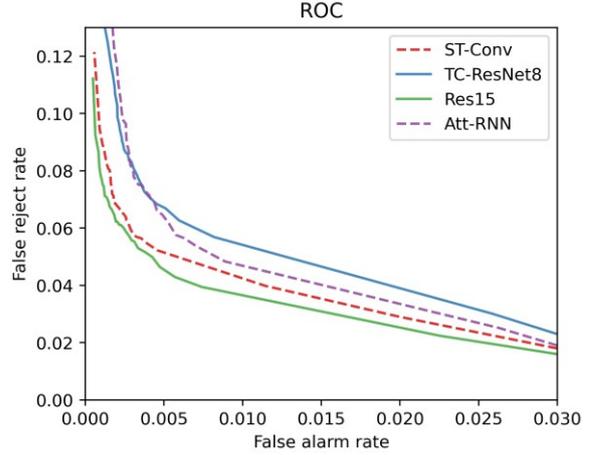

Figure 3: *ROC curves for different models.*

Compared with Res15, Att-RNN greatly reduces the number of multipliers while maintaining close accuracy. This advantage can accelerate the processing speed of the network. Compared with Att-RNN, TC-ResNet14 further reduces the number of parameters and the number of multipliers, while maintaining a consistent accuracy rate. Finally, in the proposed ST-Conv, when the number of parameters is a quarter of TC-ResNet14, the accuracy is close to TC-ResNet14 and the number of multipliers is slightly less than TC-ResNet14.

We selected comparable models in Table 3 and plotted the ROC curve in Figure 3. The remaining models have been omitted for clarity. All curves are drawn using the best model in the testing process. The smaller the area under the curve (AUC), the better the model. The results show that the performance of ST-Conv is close to Att-RNN and Res15, and better than TC-ResNet8.

## 4. Conclusions

In this paper, we introduced a separable temporal convolution model with attention for small-footprint keyword spotting. We conducted a quantitative analysis of the current state-of-the-art model and the proposed model on the publicly available Speech Commands V1 dataset. Compared with the current state-of-the-art model (TC-ResNet14), our proposed model reduces the number of parameters by two-thirds while maintaining a consistent accuracy rate. In addition, by controlling the variables, we proved that SWSA is indeed the reason for the improved accuracy of the proposed model. In our future work, we will further improve ST-Conv to make it better applicable to streaming keyword spotting.

## 5. Acknowledgements

This work is supported by National Nature Science Foundation of China (Grant No.62071039 and 61620106002).